\documentclass[prb,aps]{revtex4}
\usepackage{graphicx}

\begin{document}

\title{The anisotropic Heisenberg chain \\
in coexisting transverse and longitudinal magnetic fields}
\author{D.V.Dmitriev}
 \email{dmitriev@deom.chph.ras.ru}
\author{V.Ya.Krivnov}
\affiliation{Joint Institute of Chemical Physics of RAS, Kosygin
str.4, 117977, Moscow, Russia.}
\date{\today}

\begin{abstract}
The one-dimensional spin-1/2 $XXZ$ model in a mixed transverse and
longitudinal magnetic field is studied. Using the specially
developed version of the mean-field approximation the
order-disorder transition induced by the magnetic field is
investigated. The ground state phase diagram is obtained. The
behavior of the model in low transverse field is studied on the
base of conformal field theory. The relevance of our results to
the observed phase transition in the quasi-one-dimensional
antiferromagnet \textrm{Cs}$_2$\textrm{CoCl}$_4$ is discussed.
\end{abstract}

\maketitle

\section{Introduction}

The effects induced by magnetic fields in low-dimensional magnets
are subjects of intensive theoretical and experimental research
\cite{Dender,YbAs,Kohgi,Affleck,Lou,Ess}. One of the striking
effects is the dependence of magnetic properties of
quasi-one-dimensional antiferromagnets with anisotropic exchange
interactions on the direction of the applied magnetic field
\cite{kenz,kufo,dutta}. The basic model of such type of magnets is
the anisotropic Heisenberg chain - so-called $XXZ$ model. It is,
therefore, important to study the dependence of the properties of
the $XXZ$ chain on the field direction. There are two studied
cases of the field direction. First of them is the $XXZ$ model in
the uniform longitudinal magnetic field. This model is exactly
solved by the Bethe ansatz \cite{Yang} and is studied in great
details. In the second case the field is applied in the transverse
direction. The $XXZ$ model in the transverse field can not be
solved exactly and various approximate methods have been used to
its study \cite{mori,hieda,XXZhx,Essler}. The behavior of the
$XXZ$ model in the symmetry-breaking transverse field is
essentially different from the case of the longitudinal field. In
particular, the transverse field induces the perpendicular
antiferromagnetic long-range order (LRO) and the ground state
quantum phase transition takes place at some critical field, where
the LRO and the gap in spectrum vanish. The phase transition of
this type has been observed in the quasi-one-dimensional
antiferromagnet $\mathrm{Cs}_{2} \mathrm{CoCl}_{4}$ \cite{kenz}.
In fact, the magnetic field can have both the longitudinal and the
transverse components. For example, the magnetic field in recent
neutron scattering experiments on $\mathrm{Cs}_{2}\mathrm{
CoCl}_{4}$ has been applied at an angle to the anisotropy axes.
From this point, it is of a particular interest to study the
ground state properties of the spin $s=\frac 12$ $XXZ$ chain in
coexisting longitudinal $H_{z}$ and transverse magnetic fields
$H_{x}$. The Hamiltonian of this model is given by
\begin{equation}
H=\sum_{n=1}^{N}(S_{n}^{x}S_{n+1}^{x}+S_{n}^{y}S_{n+1}^{y}+\Delta
S_{n}^{z}S_{n+1}^{z})-h_{z}\sum_{n=1}^{N}S_{n}^{z}-h_{x}%
\sum_{n=1}^{N}S_{n}^{x}  \label{H}
\end{equation}
where
\begin{equation}
h_{x(z)}=\frac{g_{x(z)}\mu _{B}H_{x(z)}}{J}  \label{hnorma}
\end{equation}
is the effective dimensionless transverse (longitudinal) magnetic
field, $J$ is the exchange constant and $\Delta $ is the
anisotropy parameter, which is assumed to be $\Delta \geq -1$.

It was proposed \cite{kenz} that low-energy properties of
$\mathrm{Cs}_{2} \mathrm{CoCl}_{4}$ in the external magnetic field
is described by the Hamiltonian (\ref{H}) with $\Delta =0.25$ and
$J=0.23$ mev.

Evidently, in the case $\Delta =1$ the behavior of the system does
not depend on the magnetic field direction and the model (\ref{H})
reduces to the isotropic Heisenberg chain in a magnetic field
$h=\sqrt{h_z^2+h_x^2}$. In the limiting case $\Delta \rightarrow
\infty $ the model (\ref{H}) reduces to the antiferromagnetic
Ising chain in a mixed longitudinal and transverse field. This
model was investigated in \cite{sen,ZZhxhz}, where it was shown
that there is a critical line in the $(h_{x},h_{z})$ plane, where
the ground state phase transition takes place. The critical
behavior in the vicinity of this transition line belongs to the
universality class of the two-dimensional Ising model.

Thus, the physics of the model (\ref{H}) is very well understood
in the case $h_{x}=0$ and is fairly good for the cases $h_{z}=0$
and $\Delta \rightarrow \infty $, but no detailed studies are
available in general case. In this paper we study the model
(\ref{H}) using the mean-field approximation, which is the
generalization of the approach developed in \cite{XXZhx} for the
case $h_{z}=0$. This method allows us to determine the transition
line with high accuracy. The behavior in low-$h_{x}$ region will
be considered using the conformal field theory method.

The paper is organized as follows. In Sec.II we consider a
qualitative physical picture of the ground state phase diagram
based on the classical approximation. In Sec.III the mean-field
approach is developed and study of the critical properties of the
model is presented. Scaling estimations of the gap and the LRO in
low-$h_{x}$ region are given in Sec.IV. The special case $\Delta
=-1$ is studied in Sec.V. In Sec.VI we discuss our results in
relation to the experimental data for
$\mathrm{Cs}_{2}\mathrm{CoCl}_{4}$.

\section{The classical approach}

In order to provide a physical picture of the phase diagram of the
model (\ref{H}) we use the classical approximation, when spins are
represented as three-dimensional vectors. The variational wave
function corresponding to the classical approximation has a form
of a simple direct product of single-site spin states
\cite{classical}
\begin{equation}
\left| \Phi _{1}\right\rangle
=(1+A_{1}S_{1}^{+})(1+A_{2}S_{2}^{+})(1+A_{1}S_{3}^{+})(1+A_{2}S_{4}^{+})%
\ldots \left| \downarrow \downarrow \downarrow \ldots \right\rangle
\label{F1class}
\end{equation}
where $A_{1}$and $A_{2}$ are variational parameters. If $A_{1}\neq
$ $A_{2}$ then the ground state is two-fold degenerated and
another ground state wave function is
\begin{equation}
\left| \Phi _{2}\right\rangle
=(1+A_{2}S_{1}^{+})(1+A_{1}S_{2}^{+})(1+A_{2}S_{3}^{+})(1+A_{1}S_{4}^{+})%
\ldots \left| \downarrow \downarrow \downarrow \ldots \right\rangle
\label{F2class}
\end{equation}

The form of the variational parameters $A_{1}$ and $A_{2}$
minimizing the energy is different in the regions $\left| \Delta
\right| <1$ and $\Delta >1$. For the case $\left| \Delta \right|
<1$ they can be chosen as \cite{XXZhx}
\begin{equation}
A_{1}=Ae^{\mathrm{i}\phi },\quad A_{2}=Ae^{-\mathrm{i}\phi }
\label{A1A2class}
\end{equation}

The ground state energy for this case calculated with $\Phi _{1}$
(or $\Phi _{2}$) is
\begin{equation}
\frac{E}{N}=\frac{A\cos 2\phi }{(1+A^{2})^{2}}+\frac{\Delta (A^{2}-1)^{2}}{%
4(1+A^{2})}-\frac{h_{x}A\cos \phi }{1+A^{2}}-\frac{h_{z}(A^{2}-1)}{1+A^{2}}
\label{Eclass}
\end{equation}

Minimizing this energy over $A$ and $\phi $ one obtains
\begin{eqnarray}
\phi &=&\cos ^{-1}\frac{h_{x}(\Delta +1)}{2\sqrt{(\Delta +1)^{2}-h_{z}^{2}}}
\nonumber \\
A &=&\sqrt{\frac{1+\Delta +h_{z}}{1+\Delta -h_{z}}}  \label{Aficlass}
\end{eqnarray}

Two-fold degenerated ground state at $\phi \neq 0$ is
characterized by a non-zero staggered magnetization along the $Y$
direction, which plays the role of the LRO parameter
\begin{equation}
\left\langle S_{n}^{y}\right\rangle =(-1)^{n}\frac{A\sin \phi }{1+A^{2}}
\label{LROclass1}
\end{equation}

For a given value of $\Delta $ the line of phase transition on
$(h_{x},h_{z}) $ plane (the transition line) is determined by the
condition $\phi =0$ and has a form
\begin{equation}
\frac{h_{x}^{2}}{4}+\frac{h_{z}^{2}}{(1+\Delta )^{2}}=1  \label{trlineclass}
\end{equation}

\begin{figure}[tbp]
\includegraphics{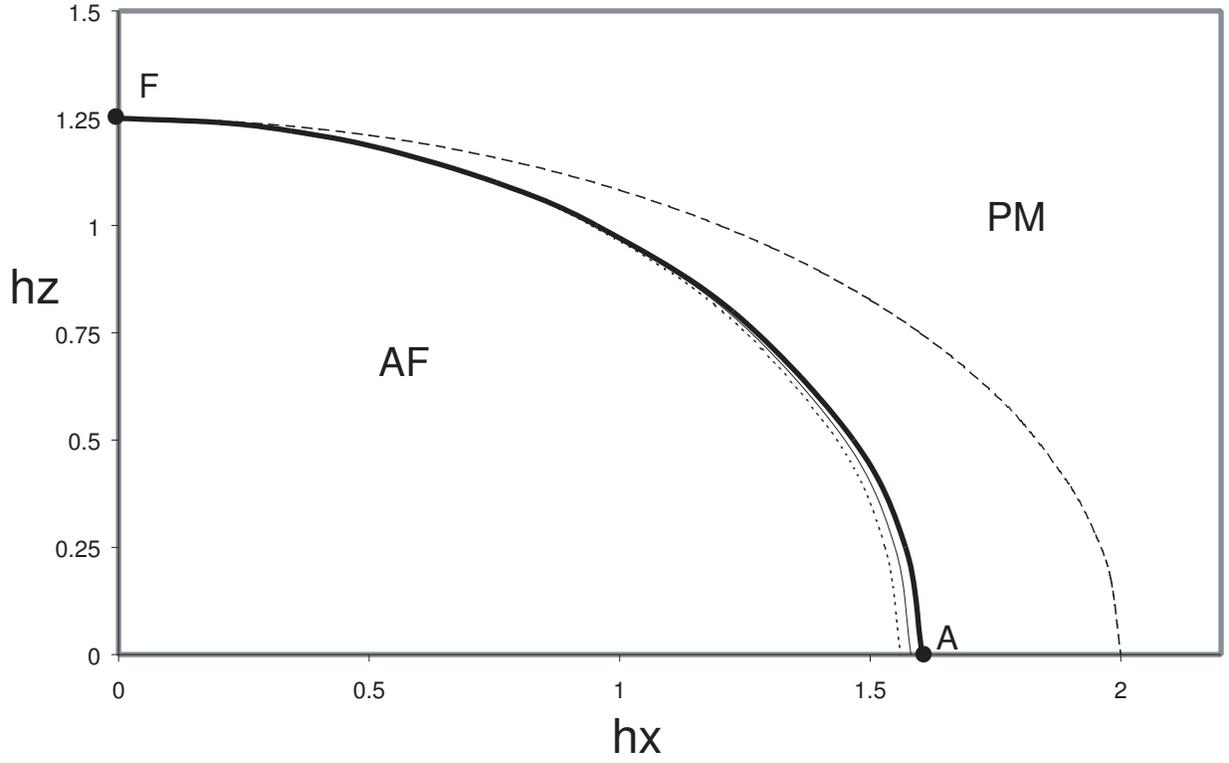}
\caption{The ground state phase diagram of the model (\ref{H}) for
$\Delta =0.25$. The transition line between the antiferromagnetic
(AF) and paramagnetic (PM) states obtained in the MFA is shown by
thick solid line and that in the classical approximation
(\ref{trlineclass}) by dashed line. Thin solid line denotes the
classical line (\ref{classline}) and dotted line corresponds to
separatrix line (see Sec.III).} \label{fig_1}
\end{figure}

This line separates the antiferromagnetic (AF) phase with the LRO
from the paramagnetic (PM) phase with uniform magnetization. The
transition line for $\Delta =0.25$ is shown on Fig.\ref{fig_1}.

\begin{figure}[t]
\includegraphics{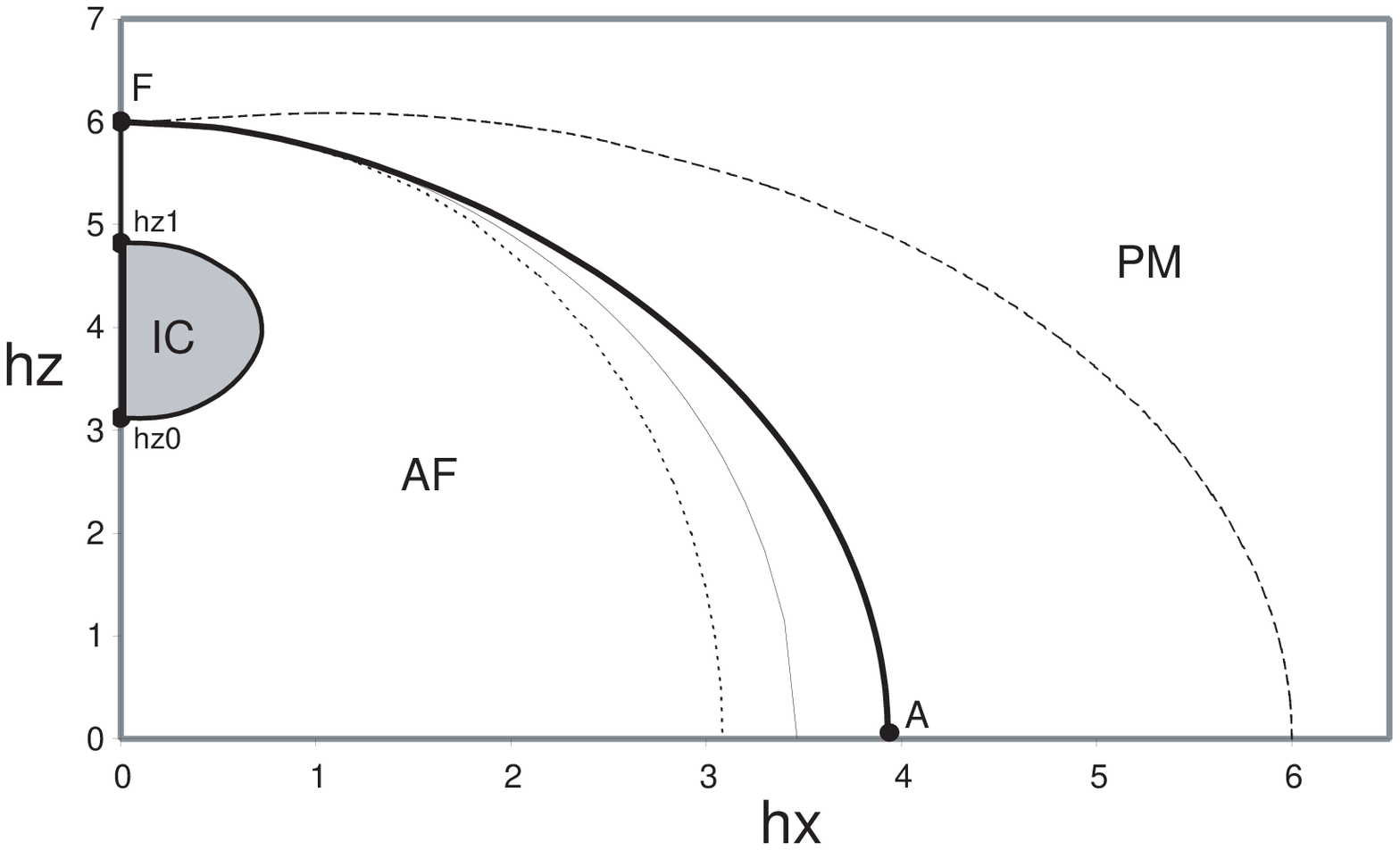}
\caption{The ground state phase diagram of the model (\ref{H}) for
$\Delta =5 $. The same notations as on Fig.1. The boundary of
incommensurate critical (IC) phase is shown schematically
(Sec.IV).} \label{fig_2}
\end{figure}

The case $\Delta >1$ can be analyzed in a similar way. In this
case two-fold degenerated ground state in the AF phase is
characterized by non-zero staggered magnetizations along the $X$
and $Z$ axes. But the expression for the transition line is rather
cumbersome and we do not present it here. The transition line in
the classical approximation for $\Delta =5$ is shown on
Fig.\ref{fig_2}.

As it is known \cite{classical} there is a remarkable, so-called
`classical' or disorder, line which lies in the AF region in the
($h_{x},h_{z}$) plane and is given by the equation:
\begin{equation}
\frac{h_{z}^{2}}{\left( 1+\Delta \right) ^{2}}+\frac{h_{x}^{2}}{2\left(
1+\Delta \right) }=1  \label{classline}
\end{equation}

The classical line is remarkable in a sense that the ground state
on it is identical to the classical one and quantum fluctuations
are missing. It was shown in \cite{classical} that the ground
state of (\ref{H}) on this line is two-fold degenerated and the
exact ground state wave functions have the product form
(\ref{F1class}) and (\ref{F2class}).

The ground state energy on the classical line for any even $N$ is
\begin{equation}
\frac{E}{N}=-\frac{1}{2}-\frac{\Delta }{4}  \label{Eclassline}
\end{equation}

From Eqs.(\ref{Aficlass}) one can find that in the case $\left|
\Delta \right| <1$ the magnetizations on the classical line are:
\begin{eqnarray}
\left\langle S_{n}^{z}\right\rangle &=&\frac{1}{2}\frac{h_{z}}{1+\Delta }
\nonumber \\
\left\langle S_{n}^{x}\right\rangle &=&\frac{h_{x}}{4}  \nonumber \\
\left\langle S_{n}^{y}\right\rangle &=&\left( -1\right) ^{n}\frac{h_{x}}{4}%
\sqrt{\frac{1-\Delta }{1+\Delta }}  \label{LROclassline1}
\end{eqnarray}

For $\Delta >1$ the parameters $A_{1}$and $A_{2}$ on the classical line are
\begin{equation}
A_{1,2}=\frac{1+\Delta +h_{z}}{h_{x}}\left( 1\pm \sqrt{\frac{\Delta -1}{%
\Delta +1}}\right)  \label{A1A2class2}
\end{equation}
and the magnetizations on two sublattices are
\begin{eqnarray}
\left\langle S_{n}^{x}\right\rangle _{1,2} &=&\frac{A_{1,2}}{A_{1,2}^{2}+1}
\nonumber \\
\left\langle S_{n}^{z}\right\rangle _{1,2} &=&\frac{1}{2}\frac{A_{1,2}^{2}-1%
}{A_{1,2}^{2}+1}  \nonumber \\
\left\langle S_{n}^{y}\right\rangle _{1,2} &=&0  \label{LROclassline2}
\end{eqnarray}

Thus, the classical approach shows that the ground state is
different in the regions with $\left| \Delta \right| <1$ and
$\Delta >1$. For $\left| \Delta \right| <1$ the classical ground
state is given by a configuration, where the spins on odd and even
sites pointing respectively at angles $\chi $ and $-\chi $ with
respect to the $XZ$ plane. For $\Delta >1$ in the ground state all
spin vectors lie in the $XZ$ plane with the spins on odd and even
sites pointing respectively at angles $\varphi _{1}$ and $\varphi
_{2}$ with respect to the $X$ axis. This means that besides
uniform magnetizations along $X$ and $Z$ axes in the AF region
there is as well the staggered magnetizations: in the $Y$
direction for $\left| \Delta \right| <1$ and in both $X$ and $Z$
directions for $\Delta >1$. \cite{ZZhxhz} These facts are
confirmed on the classical line, where the classical approximation
gives exact ground state.

Of course, one can not expect that the classical approach gives
accurate estimation of the transition line and correct description
of the phase transition (critical exponents) \cite{XXZhx}.
Nevertheless, as it will be shown below the fact of the generation
of the staggered magnetizations in the $Y$ direction for $\left|
\Delta \right| <1$ and in both the $X$ and the $Z$ directions for
$\Delta >1$ is qualitative true.

\section{Mean-field approximation}

Previously, the mean-field approximation (MFA) has been proposed
to study the anisotropic Heisenberg chain in the transverse
magnetic field \cite{XXZhx,Essler}. It has been established that
the MFA works very well if the transverse field is sufficiently
strong and it gives qualitative results for intermediate fields.
For $\Delta >-0.5$ the MFA allows to determine with high accuracy
the critical transverse field at which the order-disorder
transition occurs and to describe correctly the behavior of the
system in the transition region. The MFA is based on the
Jordan-Wigner transformation of spin-$1/2$ operators to the Fermi
operators with the subsequent mean-field treatment of the Fermi
Hamiltonian. In the case of coexisting transverse and longitudinal
magnetic fields it is impossible to reduce the model Hamiltonian
(\ref{H}) to a local form in terms of the Fermi operators.
Nevertheless, for this complicated case the MFA can be modified.
In this section we develop the special version of the MFA, which
remains the variational approach. This approach gives high
accuracy in determining of the transition line and correctly
describes the whole ground state phase diagram.

At first we perform a rotation of the spins in the $XZ$ plane by
an angle $\varphi $:
\begin{eqnarray}
S_{n}^{x} &=&\sigma _{n}^{x}\cos \varphi +\sigma _{n}^{z}\sin \varphi
\nonumber \\
S_{n}^{z} &=&-\sigma _{n}^{x}\sin \varphi +\sigma _{n}^{z}\cos \varphi
\nonumber \\
S_{n}^{y} &=&\sigma _{n}^{y}  \label{Ssigma}
\end{eqnarray}
where $\sigma _{n}^{\alpha }$ are new spin-$1/2$ operators.

The Hamiltonian (\ref{H}) is transformed to the form
\begin{eqnarray}
H &=&\sum \left( x\sigma _{n}^{x}\sigma _{n+1}^{x}+\sigma _{n}^{y}\sigma
_{n+1}^{y}+z\sigma _{n}^{z}\sigma _{n+1}^{z}\right) -h\sum \sigma
_{n}^{z}+H^{\prime }  \nonumber \\
H^{\prime } &=&\frac{1-\Delta }{2}\sin 2\varphi \sum \left( \sigma
_{n}^{x}\sigma _{n+1}^{z}+\sigma _{n}^{z}\sigma _{n+1}^{x}\right) -\left(
h_{x}\cos \varphi -h_{z}\sin \varphi \right) \sum \sigma _{n}^{x}
\label{Hrot}
\end{eqnarray}
where
\begin{eqnarray}
x &=&\cos ^{2}\varphi +\Delta \sin ^{2}\varphi  \nonumber \\
z &=&\Delta \cos ^{2}\varphi +\sin ^{2}\varphi  \nonumber \\
h &=&h_{z}\cos \varphi +h_{x}\sin \varphi  \label{xyzh}
\end{eqnarray}

The angle $\varphi $ is a variational parameter over which we will
minimize the ground state energy.

After Jordan-Wigner transformation to the Fermi operators
$a_{n}^{+}$, $a_{n} $
\begin{eqnarray}
\sigma _{n}^{+} &=&e^{i\pi \sum_{j<n}a_{j}^{+}a_{j}}a_{n}  \nonumber \\
\sigma _{n}^{z} &=&\frac{1}{2}-a_{n}^{+}a_{n}  \label{JW}
\end{eqnarray}
the Hamiltonian (\ref{Hrot}) takes the form
\begin{eqnarray}
H_{\mathrm{f}} &=&-\frac{hN}{2}+\frac{zN}{4}+\sum (h-z+\frac{1+x}{2}\cos
k)a_{k}^{+}a_{k}  \nonumber \\
&+&\frac{1-x}{4}\sum \sin k(a_{k}^{+}a_{-k}^{+}+a_{-k}a_{k})+z\sum
a_{n}^{+}a_{n}a_{n+1}^{+}a_{n+1}+H_{\mathrm{f}}^{\prime }  \label{Hfermi}
\end{eqnarray}

We treat the Hamiltonian $H_{\mathrm{f}}$ in the MFA, which
implies the decoupling of the four fermion term. The Fermi
representation $H_{\mathrm{f}}^{\prime }$ has non-local form. But
we note, that all terms in $H_{\mathrm{f}}^{\prime }$ contain odd
number of the Fermi operators $a_{n}$ and, therefore, $\langle
H_{\mathrm{f}}^{\prime }\rangle =0$ in the MFA. This fact holds
the MFA in the frame of variational principle.

Thus, in the MFA the ground state energy $E_{0}$ and the
one-particle excitation spectrum $\varepsilon (k)$ have the form:
\begin{eqnarray}
E_{0}/N &=&-\frac{h}{2}+\frac{z}{4}+(h-z)\gamma _{1}+\frac{1+x}{2}\gamma
_{2}+\frac{1-x}{4}\gamma _{3}+z\left( \gamma _{1}^{2}-\gamma _{2}^{2}+\gamma
_{3}^{2}\right)  \label{Efermi} \\
\varepsilon (k) &=&\sqrt{\left( u+v\cos k\right) ^{2}+w^{2}\sin ^{2}k}
\label{spectrum}
\end{eqnarray}
where
\begin{eqnarray}
u &=&h-z+2z\gamma _{1}  \nonumber \\
v &=&\frac{1+x}{2}-2z\gamma _{2}  \nonumber \\
w &=&\frac{1-x}{2}+2z\gamma _{3}  \label{uvw}
\end{eqnarray}

Quantities $\gamma _{1}$, $\gamma _{2}$ and $\gamma _{3}$ are the
ground state expectation values, which are determined by the
self-consistent equations:
\begin{eqnarray}
\gamma _{1} &=&\langle a_{n}^{+}a_{n}\rangle =\int\limits_{0}^{\pi }\frac{%
\mathrm{d}k}{2\pi }\left( 1-\frac{u+v\cos k}{\varepsilon (k)}\right)
\nonumber \\
\gamma _{2} &=&\langle a_{n}^{+}a_{n+1}\rangle =-\int\limits_{0}^{\pi }\frac{%
\mathrm{d}k}{2\pi }\frac{\left( u+v\cos k\right) \cos k}{\varepsilon (k)}
\nonumber \\
\gamma _{3} &=&\langle a_{n}^{+}a_{n+1}^{+}\rangle =-\int\limits_{0}^{\pi }%
\frac{\mathrm{d}k}{2\pi }\frac{w\sin ^{2}k}{\varepsilon (k)}  \label{gamma}
\end{eqnarray}

The solution of the self-consistent equations (\ref{gamma}) gives
the minimum of the ground state energy (\ref{Efermi}) in a class
of a `one-particle' wave functions at a given angle $\varphi $.
Thus, one should minimize the energy (\ref{Efermi}) with respect
to the angle $\varphi $, solving the self-consistent equations
(\ref{gamma}) for each value of $\varphi $. This means that the
proposed procedure remains variational one.

The physical meaning of the angle $\varphi $ is to show a
direction of the total magnetization of the model (\ref{H})
\begin{eqnarray}
S^{z} &=&\left\langle \sigma _{n}^{z}\right\rangle \cos \varphi =\left(
\frac{1}{2}-\gamma _{1}\right) \cos \varphi  \nonumber \\
S^{x} &=&\left\langle \sigma _{n}^{z}\right\rangle \sin \varphi =\left(
\frac{1}{2}-\gamma _{1}\right) \sin \varphi  \label{magnMF}
\end{eqnarray}

Transforming the mean-field treated Fermi Hamiltonian back to the
spin operators, we arrive to the well-studied anisotropic $XY$
model in a longitudinal magnetic field \cite{McCoy}
\begin{equation}
H_{XY}=\sum \left[ (v-w)\sigma _{n}^{x}\sigma _{n+1}^{x}+(v+w)\sigma
_{n}^{y}\sigma _{n+1}^{y}\right] -u\sum \sigma _{n}^{z}  \label{HXY}
\end{equation}

The model (\ref{HXY}) has a transition line defined by the equation
\begin{equation}
u(h_{x},h_{z},\Delta )=v(h_{x},h_{z},\Delta )  \label{translineMF}
\end{equation}
which separates the region $u<v$ with the LRO represented by a
staggered magnetization from the region $u>v$, where there is no
LRO except the uniform magnetization (\ref{magnMF}). The
transition line $h_{z\mathrm{c}}(h_{x},\Delta )$ is determined by
the numerical solution of Eqs.(\ref{gamma}),(\ref{translineMF})
with the minimization of the ground state energy over angle
$\varphi $. The transition lines in the MFA for $\Delta =0.25$ and
$\Delta =5$ are shown on Fig.\ref{fig_1} and Fig.\ref{fig_2} by
thick solid lines.

It is well known \cite{McCoy} that the critical properties of the
model (\ref {HXY}) belongs to the universality class of the
two-dimensional Ising model. This means that in the MFA the gap is
closed near the transition line linearly with the field and by the
$1/8$ law for the staggered magnetization.

The MFA also shows, that for $\left| \Delta \right| <1$ ($w>0$)
the model has a staggered magnetization along the $Y$ axis
\begin{equation}
\left\langle (-1)^{n}S_{n}^{y}\right\rangle =\frac{\left[ w^{2}\left(
v^{2}-u^{2}\right) \right] ^{1/8}}{\sqrt{2\left( v+w\right) }}  \label{SyMF}
\end{equation}
while for $\Delta >1$ ($w<0$) the staggered magnetizations exist
along the $X$ and the $Z$ axes
\begin{eqnarray}
\ \left\langle (-1)^{n}S_{n}^{x}\right\rangle &=&\frac{\left[ w^{2}\left(
v^{2}-u^{2}\right) \right] ^{1/8}}{\sqrt{2\left( v-w\right) }}\cos \varphi
\nonumber \\
\ \left\langle (-1)^{n}S_{n}^{z}\right\rangle &=&\frac{\left[ w^{2}\left(
v^{2}-u^{2}\right) \right] ^{1/8}}{\sqrt{2\left( v-w\right) }}\sin \varphi
\label{SxSzMF}
\end{eqnarray}
\qquad

\begin{figure}[tbp]
\includegraphics{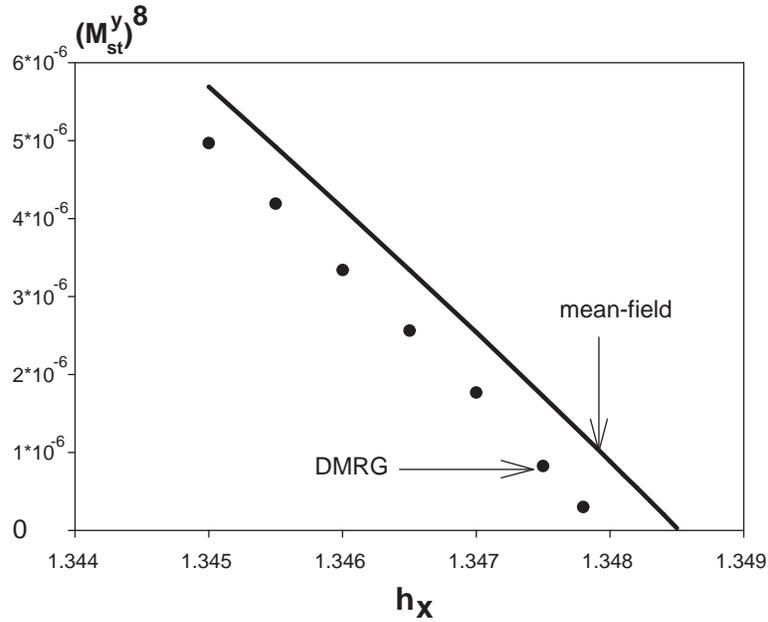}
\caption{The staggered magnetization near the transition line as a
function of $h_{x}$ for $\Delta =0.25$ and $h_{z}=0.67$. }
\label{fig_3}
\end{figure}

\begin{figure}[t]
\includegraphics{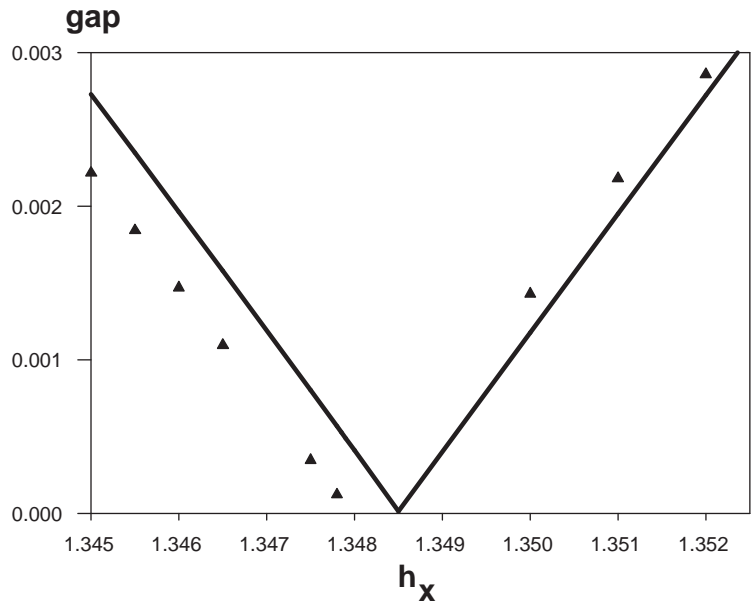}
\caption{The gap as a function of $h_{x}$ near the transition line
for $\Delta =0.25$ and $h_{z}=0.67$. (Solid line is the MFA,
triangles are DMRG results extrapolated to the thermodynamic
limit).} \label{fig_4}
\end{figure}

The validity of the 2D Ising type of the critical properties of
the model (\ref{H}) in the vicinity of the transition line has
been checked by the density matrix renormalization group (DMRG)
\cite{white} calculations of the staggered magnetization and the
gap. The staggered magnetization is computed as \cite{Essler}
\begin{equation}
M_{st}^{\alpha }=\frac{1}{N}\left\langle 0\left| \sum (-1)^{n}S_{n}^{\alpha
}\right| 1\right\rangle \qquad \alpha =(x,y,z)  \label{Mst}
\end{equation}
where $\left| 0\right\rangle $ and $\left| 1\right\rangle $ are
two lowest energy states. These states are degenerate (at
$N\longrightarrow \infty )$ in the ordered AF phase. Therefore, in
the AF phase the gap is given by the second excited state, while
in the disordered PM phase the ground state is non-degenerate and
the first excitation determines the gap in the spectrum. We have
performed the DMRG calculations using the infinite-size algorithm
and open boundary conditions and the number of states $s$ kept in
the DMRG truncating procedure is up to 25. We estimated the
relative error due to DMRG truncation from difference between the
data computed with $s=25$ and those with $s=20$ for chain lengths
$N=202$. The estimated relative error is of the order of $10^{-5}$
which is sufficiently small for accurate estimates for the gap and
the staggered magnetization. As an example, on Figs.\ref{fig_3}
and \ref {fig_4} we show the plots of $(M_{st}^{y})^{8}$ and the
gap $m$ versus $h_{x} $ in the vicinity of the transition point
for $\Delta =0.25$ and fixed $h_{z}=0.67$ (these parameters are
related to those for the antiferromagnet
$\mathrm{Cs}_{2}\mathrm{CoCl}_{4}$). A good linearity of the
plotted data definitely confirms the 2D Ising character of the
transition line. The excellent agreement between the DMRG and the
MFA results on Fig.\ref{fig_3} and Fig.\ref{fig_4} shows high
accuracy of the MFA. For example, the critical field $h_{x}$
estimated from the DMRG results differs from that obtained in the
MFA within 0.04\%.

In addition to the transition line defined by
Eq.(\ref{translineMF}), the Hamiltonian (\ref{HXY}) contains
another special line defined by the equation
\begin{equation}
u^{2}+w^{2}=v^{2}  \label{classlineMF}
\end{equation}

This line separates the so-called `oscillatory' region
$u^{2}+w^{2}<v^{2}$ (lying totally in the AF phase), where spin
correlators of the model (\ref {HXY}) have oscillatory behavior
with an incommensurate wavelength depending on the model
parameters ($h_{x},h_{z},\Delta $), from the region without such
oscillatory behavior of correlators \cite{McCoy}. The line (\ref
{classlineMF}) is nothing but the classical line of the model
(\ref{H}). Remarkably, the MFA gives the exact ground state on the
classical line. On this line the solution of Eqs.(\ref{gamma}) has
a simple form:
\begin{equation}
\sin ^{2}\varphi =\frac{p}{1-p}\frac{1+\Delta }{1-\Delta }  \label{phi_class}
\end{equation}
where
\begin{equation}
p=\frac{h_{x}^{2}}{4}\frac{1-\Delta }{1+\Delta }  \label{p}
\end{equation}

The values of $\gamma _{i}$ for $\left| \Delta \right| <1$ are
\begin{equation}
\gamma _{1}=\frac{1}{2}-\frac{\sqrt{1-p}}{2},\qquad \gamma _{2}=\gamma _{3}=-%
\frac{p}{4}  \label{gamma1}
\end{equation}
and for $\Delta >1$
\begin{equation}
\gamma _{1}=\frac{1}{2}-\frac{1}{2\sqrt{1-p}},\qquad \gamma _{2}=-\gamma
_{3}=\frac{p}{4\left( 1-p\right) }  \label{gamma2}
\end{equation}

The ground state energy is given by (\ref{Eclassline}).
Substituting Eqs.(\ref{phi_class})-(\ref{gamma2}) into
Eqs.(\ref{magnMF}),(\ref{SyMF}),(\ref {SxSzMF}) one can check that
the magnetizations on the classical line in the MFA coincide with
those given by Eqs.(\ref{LROclassline1}),(\ref {LROclassline2}).

The gap on the classical line in the MFA lies at $k=\pi $ and
equals
\begin{eqnarray}
m &=&1-\frac{p}{2}-\sqrt{1-p},\qquad \left| \Delta \right| <1  \nonumber \\
m &=&\left( 1-\frac{p}{2}-\sqrt{1-p}\right) \frac{1-2p-p\Delta }{\left(
1-p\right) ^{2}},\qquad \Delta >1  \label{mclass}
\end{eqnarray}

It is necessary to note that the elementary excitation in the AF
phase can be regarded as a domain wall between the two AF ground
states. In the cyclic chain these excitations are created in pairs
in contrast to open chains. In Eq.(\ref{Hfermi}) the end-chain
correction term is omitted, and the spectrum Eq.(\ref{spectrum})
determines the gap for the open chain \cite{LSM,Pfeuty}.
Therefore, in the AF region Eqs.(\ref{spectrum}),(\ref{mclass})
give a half of the gap for a cyclic chain.

It is worth to mention one more special line on the phase diagram,
so-called `separatrix', defined by the equation
\begin{equation}
uv=v^{2}-w^{2}  \label{separatrix}
\end{equation}

This line separates the region $uv>v^{2}-w^{2}$, where the lowest
excitation has momentum $k_{\min }=\pi $ from the region
$uv<v^{2}-w^{2}$ (situated entirely in the AF phase), where the
lowest excitation has momentum, depending on the model parameters
as
\[
\cos k_{\min }=\frac{uv}{w^{2}-v^{2}}
\]

For any $\Delta $, the transition line
$h_{z\mathrm{c}}(h_{x},\Delta )$, the classical line
$h_{z\mathrm{cl}}(h_{x},\Delta )$ and the separatrix
$h_{z\mathrm{s}}(h_{x},\Delta )$ lie in the following sequence
(see Fig.\ref {fig_1} and Fig.\ref{fig_2})
\begin{equation}
h_{z\mathrm{s}}(h_{x},\Delta )\leq h_{z\mathrm{cl}}(h_{x},\Delta
)\leq h_{z\mathrm{c}}(h_{x},\Delta )  \label{seqline}
\end{equation}

All these lines meet each other in the only point \textrm{F}
($h_{x}=0,h_{z}=1+\Delta $).

\subsection{The point \textrm{F}}

The point \textrm{F }($h_{x}=0$, $h_{z}=1+\Delta $) is the special
boundary point, where all special lines terminate. The ground
state in the point \textrm{F} is saturated ferromagnet. Near the
point \textrm{F} ($h_{x}\ll 1$) the fermion density is small and
the mean-field treatment of the four fermion term in
Eq.(\ref{Hfermi}) gives the accuracy, at least, up to $h_{x}^{4}$.
We omit here intermediate calculations and give the final
expressions for the special lines. The transition and the
separatrix lines near the point \textrm{F} have the form:
\begin{eqnarray}
h_{z\mathrm{c}}(h_{x},\Delta ) &=&h_{z\mathrm{cl}}(h_{x},\Delta
)+m_{\mathrm{cl}}+O\left( h_{x}^{6}\right)  \label{trlineF} \\
h_{z\mathrm{s}}(h_{x},\Delta ) &=&h_{z\mathrm{cl}}(h_{x},\Delta
)-m_{\mathrm{cl}}+O\left( h_{x}^{6}\right)  \label{separatrixF}
\end{eqnarray}
where the behavior of the classical line
$h_{z\mathrm{cl}}(h_{x},\Delta )$ is given by Eq.(\ref{classline})
\begin{equation}
h_{z\mathrm{cl}}(h_{x},\Delta )=1+\Delta -\frac{h_{x}^{2}}{4}-\frac{h_{x}^{4}%
}{32\left( 1+\Delta \right) }+O\left( h_{x}^{6}\right)  \label{classlineF}
\end{equation}
and $m_{\mathrm{cl}}$ is the gap near the point \textrm{F} on the
classical line (see Eq.(\ref{mclass}))
\begin{equation}
m_{\mathrm{cl}}=\frac{h_{x}^{4}}{128}\left( \frac{1-\Delta }{1+\Delta }%
\right) ^{2}+O\left( h_{x}^{6}\right)  \label{mclassF}
\end{equation}

As one can see the difference between three special lines near the
point \textrm{F} is very small, of the order of $h_{x}^{4}$.

The expressions for the gap are different to the left and to right
of the separatrix line:
\begin{eqnarray}
m &=&\frac{h_{x}^{2}}{4\sqrt{2}}\left| \frac{1-\Delta }{1+\Delta }\right|
\sqrt{h_{z\mathrm{cl}}(h_{x},\Delta )-h_{z}},\qquad h_{z}<h_{z\mathrm{s}}
\nonumber \\
m &=&\left| h_{z}-h_{z\mathrm{c}}(h_{x},\Delta )\right| ,\qquad h_{z}>h_{z%
\mathrm{s}}  \label{mF}
\end{eqnarray}

The linear behavior of the gap in the vicinity of the transition line
confirms the 2D Ising universality class of the transition line.

The staggered magnetizations in the vicinity of the point
\textrm{F} vanish on two lines: on the transition line and on the
line $h_{x}=0$. The Eqs.(\ref {magnMF}), (\ref{SyMF}),
(\ref{SxSzMF}) near the point \textrm{F} reduce to
\begin{eqnarray}
\left\langle S_{n}^{y}\right\rangle &=&(-1)^{n}B  \label{SyF} \\
B &=&\frac{1}{2}\left| \frac{1-\Delta }{1+\Delta }\right| ^{1/4}\sqrt{h_{x}}%
\left( \frac{h_{z\mathrm{c}}-h_{z}}{2}\right) ^{1/8}  \nonumber
\end{eqnarray}
for $\left| \Delta \right| <1$ and
\begin{eqnarray}
\left\langle S_{n}^{x}\right\rangle &=&\frac{h_{x}}{4}+(-1)^{n}B  \nonumber
\\
\left\langle S_{n}^{z}\right\rangle &=&\frac{1}{2}-\frac{(-1)^{n}}{2}h_{x}B
\label{SxSzF}
\end{eqnarray}
for $\Delta >1$.

To validate our analysis in the vicinity of the point \textrm{F}
one should also estimate the effect of the part of the Hamiltonian
$H^{\prime }$ in Eq.(\ref{Hrot}), omitted in the MFA. Near the
point \textrm{F} the angle $\varphi \approx h_{x}/2$ and,
therefore, these terms in $H^{\prime }$ are small and can be
taking into account as perturbations. The corresponding
perturbation theory contains only even orders. The estimate of the
second order shows that the contribution of these terms to the
ground state energy and to the gap is of the order of $h_{x}^{6}$
and $h_{x}^{2}(h_{z\mathrm{c}}-h_{z})$. This accuracy is
sufficient to confirm the above equations.

We note that in the limit $\Delta \rightarrow \infty $ the point
\textrm{F} transforms to the so-called multicritical point with
macroscopic degeneracy of the ground state \cite{domb}.

\subsection{The point \textrm{A}}

In the case $h_z=0$ the model (\ref{H}) reduces to the anisotropic
Heisenberg chain in the transverse magnetic field, which was
studied in \cite{mori,XXZhx,Essler}. At some value of magnetic
field $h_{x \mathrm{A}}(\Delta )$ this model undergoes the
transition from the antiferromagnetic state to the paramagnetic
gapful state. We denote this transition point by the point
\textrm{A} (see Figs.\ref{fig_1} and \ref {fig_2}).

To study the behavior of the system in the vicinity of the point
\textrm{A} we shall follow the arguments of \cite{ZZhxhz}, where
the point \textrm{A} was analyzed in detail for the special case
$\Delta \rightarrow \infty $. For $h_{x}=h_{x\mathrm{A}}(\Delta )$
and small longitudinal magnetic field $h_{z}$ we rewrite the
Hamiltonian (\ref{H}) in the form
\begin{eqnarray}
H &=&H_{0}+V  \nonumber \\
H_{0} &=&\sum (S_{n}^{x}S_{n+1}^{x}+S_{n}^{y}S_{n+1}^{y}+\Delta
S_{n}^{z}S_{n+1}^{z})-h_{x\mathrm{A}}(\Delta )\sum S_{n}^{x}  \nonumber \\
V &=&-h_{z}\sum S_{n}^{z}  \label{HA}
\end{eqnarray}

The unperturbed Hamiltonian $H_{0}$ describes the transition point
\textrm{A}, where the spectrum is gapless \cite{XXZhx}.

As was mentioned above, the MFA does not give the transition point
$h_{x\mathrm{A}}(\Delta )$ exactly (though with high accuracy),
but it correctly describes the character of the transition.
Therefore, we can use the MFA to determine how the small
perturbation $V$ opens the gap. [We note, that the minimum of the
MFA ground state energy at the point \textrm{A} is achieved at the
angle $\varphi =\pi /2$ and two `bad' terms in $H^{\prime }$ in
Eq.(\ref{Hrot}) disappear at this point.]

To study the perturbation theory with small perturbation $V$ one
can repeat step by step the analysis done for the special case
$\Delta \rightarrow \infty $ of the model (\ref{HA})
\cite{ZZhxhz}. As a result one finds that the perturbation theory
contains infrared divergencies, which are absorbed in the scaling
parameter $y=h_{z}^{2}N$. The perturbation series for the mass gap
has the form
\begin{equation}
m=a(\Delta )h_{z}^{2}+g\left( \Delta ,y\right) h_{z}^{2}
\label{mserA}
\end{equation}
where $a(\Delta )$ is the second order correction and $g\left(
\Delta ,y\right) $ is the scaling function accumulated all
high-order divergent terms. The scaling function $g\left( \Delta
,y\right) $ in the thermodynamic limit ($y\rightarrow \infty $)
tends to some finite limit. So, the gap takes the form
\begin{equation}
m=\left[ a(\Delta )+g\left( \Delta ,\infty \right) \right]
h_{z}^{2} \label{mA1}
\end{equation}

From the last equation we see that the gap is proportional to
$h_{z}^{2}$, but the factor at $h_{z}^{2}$ is given not only by
the second order correction $a(\Delta )$ but by all collected
divergent orders of the perturbation series.

For a fixed value of $\Delta $ the behavior of the transition line
in ($h_{x},h_{z}$) plane near the point \textrm{A} can be found
from the following consideration. As it was established above in
the vicinity of the transition line the gap is proportional to the
deviation from the line. This is valid for any direction of
deviation except the direction at a tangent to the transition
line. Thus, in the vicinity of the point \textrm{A} on the line
$h_{x}=h_{x\mathrm{A}}(\Delta )$, the gap is
\begin{equation}
m\sim h_{x\mathrm{A}}(\Delta )-h_{x\mathrm{c}}(h_{z},\Delta )
\label{mA2}
\end{equation}

On the other hand, the gap is given by Eq.(\ref{mA1}). Equalizing
these two expressions for the gap we obtain the equation for the
transition line in the vicinity of the point \textrm{A} as
\begin{equation}
h_{x\mathrm{c}}(h_{z},\Delta )=h_{x\mathrm{A}}(\Delta )-f(\Delta
)h_{z}^{2} \label{trlineA}
\end{equation}
where the function $f(\Delta )$ is generally unknown and can be
found numerically only.

Summarizing all above, we conclude that the MFA correctly
describes the critical properties of the transition line and
determines the transition line with high accuracy. This is because
the MFA gives the exact ground state on the classical line, which
is close to the transition line. Besides, the MFA is
asymptotically exact in the vicinity of the point \textrm{F}.
Therefore, for any value of $\Delta $, the accuracy in determining
of the transition line drops as one moves from the point
\textrm{F} to the point \textrm{A}. The MFA quality for the case
$h_{z}=0$ was investigated in \cite{XXZhx,Essler}, where it was
shown that the accuracy of the MFA is high for $\Delta >-0.5$, and
the MFA fails in the limit $\Delta \rightarrow -1$ (where the
accuracy decreases to 20\%). Besides, the high accuracy of the MFA
in the vicinity of the transition line is confirmed by DMRG
calculations (see Fig.\ref{fig_3} and Fig.\ref{fig_4}). The MFA
qualitative correctly describes the line $h_{x}=0$ (no gap and
sound-like spectrum for $h_{z}<1+\Delta $), but one can not expect
that the MFA gives correct critical exponents in low-$h_{x}$
region.

\section{The low-$h_x$ region}

\begin{figure}[tbp]
\includegraphics{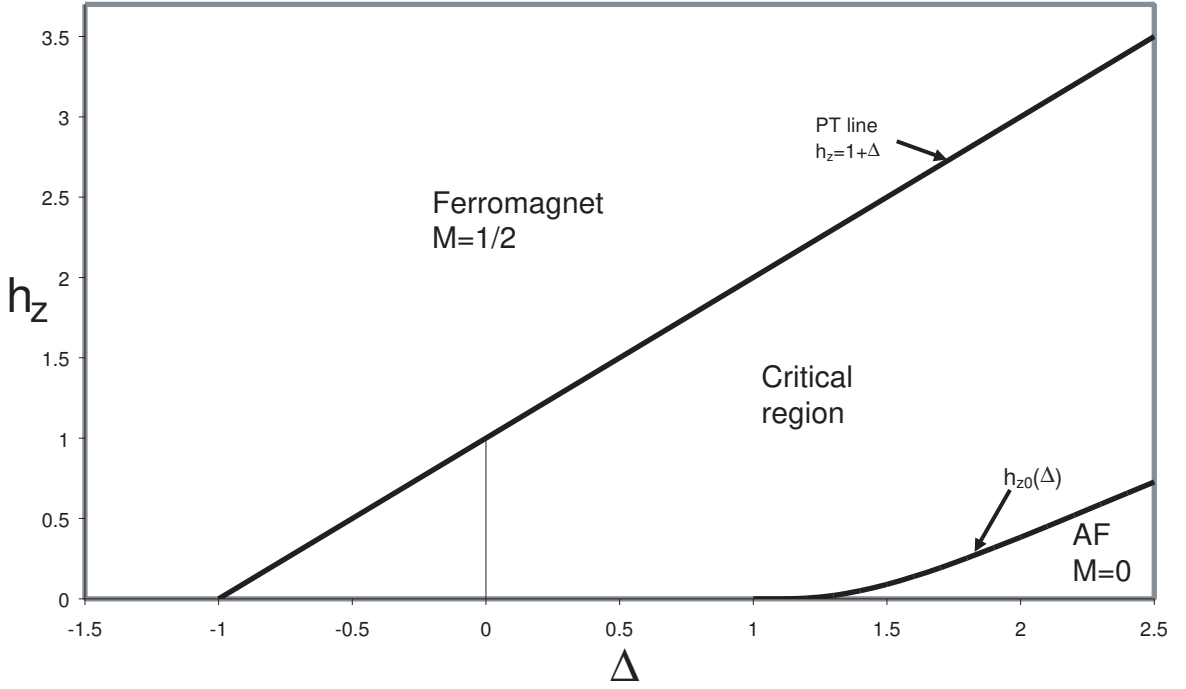}
\caption{The ground state phase diagram of the $XXZ$ model in
longitudinal magnetic field.} \label{fig_5}
\end{figure}

On the line $h_x=0$ the model (\ref{H}) reduces to the well-known
exactly solvable $XXZ$ model in the longitudinal magnetic field.
In this model three phases exist in different ranges of the
magnetic field $h_z$: the ferromagnetic (F) phase at
$h_z>1+\Delta$; the antiferromagnetic (AF) phase at
$0<h_z<h_{z0}(\Delta )$ ($h_{z0}(\Delta )$ is a lower critical
field \cite{Gaudin}) and the critical phase at $0<h_z<1+\Delta$
($|\Delta|<1$) and $h_{z0}(\Delta ) <h_z<1+\Delta$ ($\Delta>1$)
(see Fig.\ref{fig_5}).

In the F phase the ground state is saturated ferromagnet $M\equiv
\left\langle S_n^z\right\rangle =1/2$ with a gap in the spectrum.
In this region the appearance of the transverse magnetic field
does not cause noticeable change in the system properties. It
results in appearance of a uniform magnetization in the $X$
direction and small decreasing of the magnetization in the $Z$
direction.

In the AF region the system is in a gapful phase with the
long-range Neel order $M_{st}^{z}$ and zero uniform magnetization
$M=0$. Due to the gap in the spectrum the effect of the $X$
component of magnetic field in the AF region can be obtained in
the frame of a regular perturbation theory in $h_{x}$. The
estimate of the first and the second orders in $h_{x}$ indicates
the appearance of the uniform magnetizations in both the $X$ and
the $Z$ directions and the staggered magnetization in the $X$
direction as
\begin{eqnarray}
\left\langle S_{n}^{x}\right\rangle &\sim &h_{x}+(-1)^{n}h_{z}h_{x}
\nonumber \\
\left\langle S_{n}^{z}\right\rangle &\sim &(-1)^{n}M_{st}^{z}+h_{z}h_{x}^{2}
\label{LRO_AF}
\end{eqnarray}

As follows from the last equations, in the case $h_{z}=0$ the
applied transverse magnetic field does not cause the uniform
magnetizations in the $Z$ direction and the staggered
magnetization in the $X$ direction \cite{XXZhx}.

The critical phase is characterized by non-zero magnetization
$0<M<1/2$ in the ground state and by the massless spectrum. The
low-energy properties in this phase are described by a free
massless boson field theory with the Hamiltonian
\begin{equation}
H_{0}=\frac{v}{2}\int \mathrm{d}x\left[ (\partial _{x}\Theta )^{2}+(\partial
_{x}\Phi )^{2}\right]  \label{Hgauss}
\end{equation}
where $\Phi (x)$ and $\Theta (x)$ are boson and dual field,
respectively and $v(\Delta,h_z)$ is the renormalized spin-wave
velocity.

The spin-density operators are represented as \cite{book}
\begin{eqnarray}
S_n^z &\simeq & M + \frac{1}{2\pi R}\partial _{x}\Phi +
a_1 (-1)^n \cos\left(\frac{\Phi}{R}+2\pi Mx\right)  \nonumber \\
S_n^x &\simeq & b_0(-1)^{n}\cos \left( 2\pi R\Theta \right)
+b_1\cos \left( 2\pi R\Theta \right)\cos \left( \frac{\Phi
}{R}+2\pi Mx\right) \label{Sbosonhz}
\end{eqnarray}
where $a_1,b_0$ and $b_1$ are some constants \cite{Hikihara} and
we identify the site index $n$ with the continuous space variable
$x$. The magnetization $M(\Delta,h_z)$ and the compactification
radius $R(\Delta,M)$ are functions of $\Delta $ and $h_z$ and can
be determined by solving Bethe-ansatz integral equations
\cite{BIK,Cabra}.

Both terms of operator $S^x$ in Eq.(\ref{Sbosonhz}) are
oscillating when $M\neq 0$ and are not relevant to the uniform
$X$-component of the magnetic field. But as was shown in
\cite{Nersesyan} the second term in Eq.(\ref{Sbosonhz})
corresponding to perturbation
\begin{equation}
V_0= h_x b_1 \cos \left( 2\pi R\Theta \right) \cos \left(
\frac{\Phi }{R}+2\pi Mx\right)  \label{V0hz}
\end{equation}
has conformal spin $S=1$ and generates two other perturbations
with zero conformal spin:
\begin{eqnarray}
V_1 &=&g_1\cos \left( 4\pi R\Theta \right)  \nonumber \\
V_2 &=&g_2\cos \left( \frac{2\Phi }{R}+4\pi Mx\right)
\label{V1V2hz}
\end{eqnarray}
with $g_1\sim h_x^2$ and $g_2\sim h_x^2$.

The scaling dimensions of perturbations $V_1$, $V_2$ are $2\eta $
and $2/\eta $ ($\eta=2\pi R^2$), respectively. The perturbation
$V_2$ describes Umklapp processes and it is responsible for the
gap generation in the AF region, where $M=0$. But in the critical
region ($h_z>0$) the magnetization $M\neq 0$ and the perturbation
$V_2$ as well as the operator $V_0$ does not conserve the total
momentum and will be frozen out.

Therefore, in the uniform longitudinal magnetic field $h_z$ the
critical exponent for the mass gap is determined by the only
non-oscillating perturbation $V_1$:
\begin{equation}
m\sim h_x^{\frac{1}{1-\eta}}  \label{mB}
\end{equation}

We note that in the special case $h_z=0$ ($|\Delta|<1$) the
magnetization $M=0$ and all perturbations $V_0$, $V_1$ and $V_2$
are non-oscillated \cite{Bocquet}. In this case in the region
$\Delta>\cos [\pi \sqrt{2}]\approx -0.266$ the perturbation $V_0$
becomes most relevant and determines the mass gap as \cite{XXZhx}
\begin{equation}
m\sim h_{x}^{\frac{2}{4-\eta -1/\eta }}  \label{malpha}
\end{equation}

The perturbation $V_1$ corresponds to the spin-nonconserving
operator $\sum(S^x_n S^x_{n+1}-S^y_n S^y_{n+1})$.\cite{book} This
means that the behavior of the system (\ref{H}) at small $h_x$ is
similar to that of well studied $XYZ$ chain in magnetic field
$h_z$ with small anisotropy in $XY$ plane \cite{Giamarchi}. Using
the results of Ref.\cite{Giamarchi} we conclude that in the region
$\Delta>1$ the presence of small $X$-component of magnetic field
leads to a sequence of three transitions with increasing
longitudinal magnetic field $h_z$. The first one occurs at
$h_{z0}(\Delta)$ ($\eta =2$), where the AF phase transforms to the
incommensurate critical (IC) or `floating' phase (see
Fig.\ref{fig_2}). In the IC phase the perturbation $V_1$ is
irrelevant and the spectrum remains gapless. Here the correlation
functions display a power-law decay with a magnetization dependent
wave vector. There is no LRO in the IC phase, except uniform
magnetizations $M$ and $\left\langle S_n^x\right\rangle=\chi_x
h_x$ (the susceptibility $\chi_x$ is finite in the critical
phase). Unfortunately, neither the field theory approach nor the
MFA allow us to determine the boundary of the IC phase. Therefore,
this phase boundary is shown on Fig.\ref{fig_2} schematically.

Further increasing of $h_z$ leads to the transition of the
Kosterlitz-Thouless type taking place at the point
$h_{z1}(\Delta)$ where $\eta=1$. At $h_z>h_{z1}(\Delta)$ the
perturbation $V_1$ becomes relevant ($\eta<1$) and the system
crosses from the IC phase to a strong-coupling regime with the
staggered magnetizations in both $X$ and $Z$ directions (AF
phase).

The last transition with further increasing of $h_z$ occurs near
the point \textrm{F} at $h_z=h_{zc}$ (see Fig.\ref{fig_2}). This
transition to the PM phase was studied in Sec.III.

The field theory approach allows also to determine the exponent
for the appeared LRO. The transverse magnetic field generates the
staggered magnetization along the $Y$ axis at
$\left|\Delta\right|<1$ as
\begin{equation}
\left\langle S_{n}^{y}\right\rangle \sim \frac{(-1)^{n}}{\xi ^{\eta /2}}\sim
(-1)^{n}m^{\eta /2}\sim (-1)^{n}h_{x}^{\eta /(2-2\eta )}  \label{SyB}
\end{equation}
and along $X$ and $Z$ axes at $\Delta >1$ and
$h_{z1}(\Delta)<h_z<1+\Delta$
\begin{eqnarray}
\left\langle S_{n}^{x}\right\rangle &\sim &\chi
_{x}h_{x}+(-1)^{n}h_{x}^{\eta /(2-2\eta )}  \nonumber \\
\left\langle S_{n}^{z}\right\rangle &\sim &M+(-1)^{n}h_{x}^{(2-\eta
)/(2-2\eta )}  \label{SxSzB}
\end{eqnarray}

We note, that in the limit $h_z\to 1+\Delta $ ($\eta \to 1/2$),
the exponents for the mass gap (\ref{mB}) and staggered
magnetizations in Eqs.(\ref{SyB}),(\ref{SxSzB}) agree with the MFA
results in Eqs.(\ref{mF}),(\ref{SyF}),(\ref{SxSzF}).

The staggered magnetization along the $Z$ axis in Eq.(\ref{SxSzB})
can be derived in the same manner as was derived the generated
perturbations $V_1$ and $V_2$ in Eqs.(\ref{V1V2hz}). According to
Eqs.(\ref{Sbosonhz}) the non-zero contribution to the first order
correction in $h_{x}$ to the staggered magnetization $\left\langle
(-1)^{n}S_{n}^{z}\right\rangle $ is given by the following terms
in spin-density operators
\begin{eqnarray}
(-1)^{n}S_{n}^{z} &\sim &\cos \left( \frac{\Phi }{R}+2\pi Mx\right)
\nonumber \\
S_{n}^{x} &\sim &\cos \left( 2\pi R\Theta \right) \cos \left( \frac{\Phi }{R}%
+2\pi Mx\right)  \label{Szst1}
\end{eqnarray}

Then the leading contribution comes from small distances of order
of ultraviolet cut-off (the lattice constant)
\begin{eqnarray}
(-1)^{x_{1}}S^{z}(z_{1}) &\sim &h_{x}\int d^{2}z_{2}e^{\mathrm{i}2\pi
M(x_{2}-x_{1})}\exp [-\mathrm{i}\frac{\Phi (z_{1})}{R}]\exp [\mathrm{i}\frac{%
\Phi (z_{2})}{R}]\cos [2\pi R\Theta (z_{2})]  \nonumber \\
&\sim &h_{x}\cos [2\pi R\Theta (z_{1})]\sim h_{x}(-1)^{x_{1}}S^{x}(z_{1})
\label{Szst2}
\end{eqnarray}

Thus, the relation between the staggered magnetization along $X$
and $Z$ axes is established
\begin{equation}
\left\langle (-1)^{n}S_{n}^{z}\right\rangle \sim h_{x}\left\langle
(-1)^{n}S_{n}^{x}\right\rangle  \label{Szst3}
\end{equation}
which results in the critical exponent for $\left\langle
(-1)^{n}S_{n}^{z}\right\rangle $ written in Eq.(\ref{SxSzB}).

\subsection{Perturbation theory. Mapping to the $XYZ$ model in the magnetic
field}

In this subsection we study the behavior of the model in the
critical phase from the perturbation theory point of view. In some
sense this subsection is complementary to the field theory
approach.

For a small $X$ component of a magnetic field the Hamiltonian
(\ref{H}) can be written in the form
\begin{eqnarray}
H &=&H_{0}+V  \nonumber \\
H_{0} &=&\sum (S_{n}^{x}S_{n+1}^{x}+S_{n}^{y}S_{n+1}^{y}+\Delta
S_{n}^{z}S_{n+1}^{z})-h_{z}\sum S_{n}^{z}  \nonumber \\
V &=&-h_{x}\sum S_{n}^{x}  \label{Hhx0}
\end{eqnarray}

As was said above in the critical region (\ref{fig_5}) the model
$H_{0}$ is in the critical regime with non-zero magnetization $M$.

The transition operator $S^{x}=\sum S_{n}^{x}$ in $V$ conserves
the momentum and changes the total $S^{z}$ by one. Therefore,
since the unperturbed Hamiltonian $H_{0}$ conserves the value of
$S^{z}$, the perturbation theory in $V$ contains only even orders.

The operator $V$ acting on the ground state or the low-lying
states produces states with `high' energies $E_{s}-E_{0}\equiv
\varepsilon _{s}\gtrsim 1$. So, the perturbation theory in $V$ has
such a `two-level' structure with alternate low-lying and
high-lying intermediate states. As a result the denominator of the
$2n$-th perturbation order contains $n$ factors with `high'
excitation energies $\varepsilon _{s}\gtrsim 1$ and ($n-1$) with
low-lying excitation $\varepsilon _{s}\sim 1/N$.

Omitting numerical factors, in all orders of perturbation theory
one can take out all `high' energies from the denominator and sum
up over all high-lying intermediate states. As a result, we arrive
at the perturbation theory with the effective perturbation
\begin{equation}
V^{\prime }=-h_{x}^{2}\sum S_{n}^{x}S_{m}^{x}  \label{Vp}
\end{equation}

We note, that the perturbation theory with the perturbation
$V^{\prime }$ coincides with the original perturbation theory
(\ref{Hhx0}) in a sense, that both perturbation series have the
same order of divergencies (or power of $N$) at each order in
$h_{x}$. But numerical factors at each order in $h_x$ can be
different.

The behavior of the ground state correlation function
$\left\langle S_{n}^{x}S_{n+r}^{x}\right\rangle $ of the model
$H_{0}$ on large distances is known \cite{Hikihara}
\begin{equation}
\left\langle S_{n}^{x}S_{n+r}^{x}\right\rangle =A_{0}(\eta )\frac{(-1)^{r}}{%
r^{\eta }}-A_{1}(\eta )\frac{\cos \left( 2\pi Mr\right) }{r^{\eta
+1/\eta }} \label{XXcorr}
\end{equation}
and, therefore, due to oscillation of the correlator $\left\langle
S_{n}^{x}S_{m}^{x}\right\rangle $ (for $M\neq 0$) the sum over $n$
and $m$ can be approximately estimated as a one half of the first
term:
\begin{equation}
\sum_{n,m}\left\langle S_{n}^{x}S_{m}^{x}\right\rangle =\frac{N}{4}%
+2\sum_{n>m}\left\langle S_{n}^{x}S_{m}^{x}\right\rangle \simeq \frac{N}{4}%
+\sum_{n}\left\langle S_{n}^{x}S_{n+1}^{x}\right\rangle
\label{Vpdiag}
\end{equation}

The last equation suggests that the perturbation $V^{\prime }$ can
be reduced to the operator $V^{\prime \prime }$
\begin{equation}
V^{\prime \prime }=-ah_{z}^{2}N-bh_{z}^{2}\sum
S_{n}^{x}S_{n+1}^{x} \label{Vpp}
\end{equation}
with some constants $a,b$. In order to verify this assumption we
compared the matrix elements of the operator $V^{\prime \prime }$
with those of $V^{\prime }$. We have found that the dependence of
the matrix elements of the operators $V^{\prime }$ and $V^{\prime
\prime }$ on $N$ is identical.

As a result of the analysis we arrive at the effective Hamiltonian
\begin{equation}
H_{\mathrm{eff}}=-aNh_{x}^{2}+\sum \left(
(1-bh_{x}^{2})S_{n}^{x}S_{n+1}^{x}+S_{n}^{y}S_{n+1}^{y}+\Delta
S_{n}^{z}S_{n+1}^{z}\right) -h_{z}\sum S_{n}^{z}  \label{Hmap}
\end{equation}

Again, the original model (\ref{Hhx0}) is equivalent to the
effective model (\ref{Hmap}) in a sense, that the perturbation
series for both models have the same order of divergencies at each
order of $h_{z}$. This fact means that the effective model has the
same critical exponents as the original model (\ref{Hhx0}).

One can see, that the effective perturbation $V^{\prime \prime }$
is nothing but the generated perturbation $V_{1}$ found in
bosonization technique Eq.(\ref{V1V2hz}) with $g_{1}=bh_{x}^{2}$.
So, the above analysis represents another way to obtain an
effective perturbation responsible for the most divergent part in
the perturbation theory.

Unfortunately, the effective model (\ref{Hmap}) is not exactly
solvable one in general case. But it can be used for checking the
bosonization expression for the mass gap (\ref{mB}) in two special
cases $h_{z}=0$ and $\Delta =0$, where the effective model reduces
to integrable $XYZ$ model and $XY$ model in magnetic field,
respectively. In these special cases the exact expressions for the
mass gap are known \cite{McCoy,Baxter}. They confirm the validity
of the critical exponent in Eq.(\ref{mB}).

\section{Special case $\Delta =-1$}

When $\Delta \rightarrow -1$ the point \textrm{F} tends to zero,
and exactly on the line $\Delta =-1$ both classical and transition
lines disappear. In the vicinity of the line $\Delta =-1$ it is
convenient to rotate the coordinate system in such a way that the
Hamiltonian (\ref{H}) takes the form
\begin{eqnarray}
H &=&H_{0}+V_{h}+V_{\Delta }  \nonumber \\
H_{0} &=&-\sum \mathbf{S}_{n}\cdot \mathbf{S}_{n+1}-h_{x}\sum
(-1)^{n}S_{n}^{z}  \nonumber \\
V_{h} &=&-h_{z}\sum S_{n}^{x}  \nonumber \\
V_{\Delta } &=&(1+\Delta )\sum S_{n}^{x}S_{n+1}^{x}  \label{Hd-1}
\end{eqnarray}

The unperturbed Hamiltonian $H_0$ describes the critical behavior
of the system at $0<h_x<h_0$, \cite{Alkaraz} where the estimate of
$h_0$ is $h_0\approx 0.53(3)$. \cite{XXZhx} The spin-density
operators of $H_0$ are related to the bosonic fields by
Eqs.(\ref{Sbosonhz}) with $M=0$, where the compactification radius
$R(h_x)$ is a function of $h_x$. This dependence is generally
unknown, but the limiting values of $\eta(h_x)=2\pi R^2(h_x)$ are
known \cite{XXZhx}
\begin{eqnarray}
\eta \left( h_{x}\right) &=&\frac{h_{x}}{\pi },\qquad
h_{x}\rightarrow 0
\nonumber \\
\eta \left( h_{x}\right) &=&\frac{1}{4},\qquad h_{x}\rightarrow h_{0}
\label{eta_d-1}
\end{eqnarray}

The perturbations $V_h$ and $V_{\Delta }$ have scaling dimensions
$\eta /2$ and $2\eta $, respectively. Therefore, according to
Eqs.(\ref{eta_d-1}) both perturbations $V_h$ and $V_{\Delta }$ are
relevant in the critical region $0<h_x<h_0$ and produce a mass
gaps
\begin{eqnarray}
m_{h} &\sim &h_{z}^{1/(2-\eta /2)}  \nonumber \\
m_{\Delta } &\sim &\left| 1+\Delta \right| ^{1/(2-2\eta )}  \label{m_d-1}
\end{eqnarray}

In particular, $m_{h}\sim \sqrt{h_{z}}$, $m_{\Delta }\sim
\sqrt{\left| 1+\Delta \right| }$ at $h_{x}\to 0$ and $m_{h}\sim
h_{z}^{8/15}$, $m_{\Delta }\sim \left| 1+\Delta \right| ^{2/3}$ at
$h_x\to h_0$.

The gap at $\Delta =-1$ and for $h_x,h_z<<1$ can be asymptotically
exactly described by the spin-wave theory \cite{XXZhx}. The
validity of the spin-wave approximation is quite natural because
the number of magnons forming the ground state is small for
$h_x,h_z<<1$. The spin-wave result for the gap is
\begin{equation}
m=\sqrt{h_{z}(h_{z}+\frac{h_{x}^{2}}{2})}  \label{SWTd-1}
\end{equation}

For $h_z<<h_x^2<<1$ Eq.(\ref{SWTd-1}) agrees with the conformal
theory result Eq.(\ref{m_d-1}) at $\eta\to 0$ and gives the
preexponential factor for the gap.

\section{Discussion and conclusion}

\begin{figure}[tbp]
\includegraphics{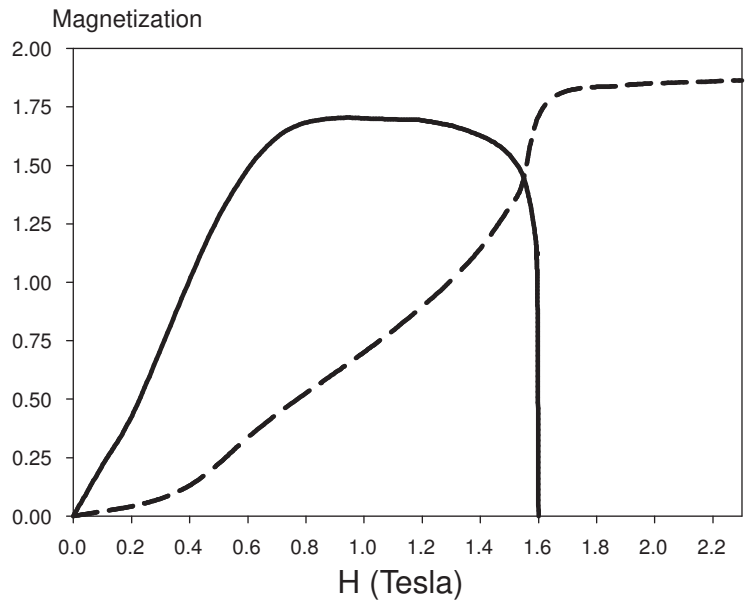}
\caption{Uniform (dashed line) and staggered (solid line)
magnetization curves.}
\label{fig_6}
\end{figure}

As it was mentioned the order-disorder transition induced by the
magnetic field has been observed in the quasi-one-dimensional
antiferromagnet $\mathrm{Cs}_{2}\mathrm{CoCl}_{4}$ described, as
supposed, by the model (\ref {H}) with $\Delta =0.25.$ It is
interesting to compare the magnetization curves obtained in
neutron-scattering experiment \cite{kenz} with those in the MFA.
In this experiment the magnetic field $H$ has been applied at an
angle $\beta \simeq 40^{0}$ to the $XY$ plane. This means that
$H_{x}\simeq H_{z}\simeq H/\sqrt{2}.$ According to \cite{kenz}
$g$-factors in $\mathrm{Cs}_{2}\mathrm{CoCl}_{4}$ are
$g_{x,y}=2g_{z}=4.80$. Therefore, the ratio of the effective
fields in the model (\ref{H}) is
$\frac{h_{x}}{h_{z}}=\frac{2H_{x}}{H_{z}}\simeq 2$. The total
magnetization $M_{tot}$ is
\begin{equation}
M_{tot}=\frac{\mu _{b}}{\sqrt{2}}\left( g_{x}\langle S_{n}^{x}\rangle
+g_{z}\langle S_{n}^{z}\rangle \right)  \label{Mexp}
\end{equation}
where $\langle S_{n}^{x}\rangle $ and $\langle S_{n}^{z}\rangle $
are the magnetizations calculated in the MFA at the effective
fields $h_{x}=2h_{z}$. The AF ordered moment is given by
$M_{st}=g_{y}\mu _{b}\langle \left| (-1)^{n}S_{n}^{y}\right|
\rangle $. On Fig.\ref{fig_6} we plot $M_{tot}$ and $M_{st}$ as
the functions of the magnetic field $H$ ($H=\frac{\sqrt{2}h_{x}J
}{g_{x}\mu _{b}})$. These magnetization curves are qualitative
similar to the experimental ones (Figs.12 and 14 in
Ref.\cite{kenz}). The maximal value of staggered magnetization
$1.7\mu _{b}$ agrees with the experimental magnitude of the AF
ordered moment 1.6$\mu _{b}$. The total magnetization at
$H\rightarrow \infty $ in the MFA $M_{tot}=1.9\mu _{b}$ is
consistent with the saturation moment $1.7\mu _{b}$ estimated in
\cite{kenz}. At the same time, there is essential difference in
the low-field behavior of $M_{st}$. The experimental AF ordered
moment is finite at $H=0$, while $M_{st}=0$ at $H=0$ on
Fig.\ref{fig_6}. This difference is due to weak interchain
couplings, which form the magnetically ordered ground state in the
real compound at $H=0$ and this effect is absent in the
one-dimensional model (\ref{H}). The behavior of the
magnetizations in the vicinity of the critical field on
Fig.\ref{fig_6} is similar to the experimental curve. But the
value of the critical field on Fig.\ref{fig_6} is $1.6T$, while
the experimental value is $2.1T$. We do not believe that the MFA
is the reason for this discrepancy. The possible reason may lie in
the fact that $\mathrm{Cs}_{2}\mathrm{CoCl}_{4}$ is described by
the $s=\frac{3}{2}$ antiferromagnetic model with strong single-ion
anisotropy and its reducing to the $s=\frac{1}{2}$ model is
approximate.

Magnetic measurements in \cite{kenz} were performed in the field
applied at a fixed angle $\beta $ to the anisotropy axis. It is
interesting to consider how the properties of the model (\ref{H})
is changed when the magnetic field is turned from purely
longitudinal ($\beta =0$) to transverse direction ($\beta
=\frac{\pi }{2}$). If the effective field $h$ is in the range
$(1+\Delta )<h<h_{x\mathrm{A}}(\Delta )$ for $\left| \Delta
\right| <1$ or $h_{x\mathrm{A}}(\Delta )<h<(1+\Delta )$ for
$\Delta >1$ then there is the critical angle $\beta _{0}$, which
is defined by the intersection of the circle
$h_{x}^{2}+h_{z}^{2}=h^{2}$ with the transition line. At this
angle the phase transition from the $AF$ phase at $\beta >\beta
_{0}$ ($\beta <\beta _{0})$ for $\Delta <1$ ($\Delta >1)$ to $PM$
phase at $\beta <\beta _{0}$ ($\beta >\beta _{0})$ takes place.
The staggered magnetization and the gap vanish at $\beta =\beta
_{0}$ as $M_{st}\sim \left| \beta -\beta _{0}\right| ^{1/8}$ and
$m\sim \left| \beta -\beta _{0}\right| $.

In conclusion, we have studied the spin-$\frac 12$ $XXZ$
Heisenberg chain in the mixed longitudinal and transverse magnetic
field. It was shown that the ground state phase diagram on the
$(h_x,h_z)$ plane contains the $AF$ and the $PM$ phases separated
by the transition line. The transition line was determined using
the proposed special version of the MFA, which reduces the $XXZ$
model in the mixed fields to the $XY$ model in the uniform
longitudinal field. The MFA gives the transition line with high
accuracy at least for $\Delta \geq -0.5$. This fact is confirmed
by comparison of the MFA results with DMRG calculations. The MFA
gives satisfactory description of the whole phase diagram, though
the critical exponents of low-$h_x$ dependence of the gap and the
magnetization can not be found correctly in the MFA. These
exponents have been found with use of the conformal field method.

The field theory approach also shows that in the region $\Delta>1$
the phase diagram contains IC or floating phase characterized by
the gapless spectrum and power-low decay of correlation functions.
The form of the boundary of IC phase can be determined numerically
only. But this boundary is located certainly on the left of the
classical line, where the spectrum is gapped.

We believe that the modified MFA is suitable for the studying of
the magnetic phase transitions of the quasi-one dimensional
anisotropic magnets induced by the applied magnetic field. The
important problem is to take into account effects of inter-chain
interactions \cite{Q1D}.

We thank Dr.V.Cheranovskii for useful discussions. This work was
supported under RFBR Grant No 03-03-32141 and ISTC No 2207.


\begin{thebibliography}{99}
\bibitem{Dender}  D.C.Dender, P.R.Hammar, D.H.Reich, C.Broholm and G.Aeppli,
Phys.Rev.Lett. 79, 1750 (1997).

\bibitem{YbAs}  R.Helfrich, M.Koppen, M.Lang, F.Steglich and A.Ochiku,
J.Magn.Magn.Mat. 177, 309 (1998).

\bibitem{Kohgi}  M.Kohgi, K.Iwasa, J.-M.Mignot, B.F\aa k, P.Gegenwart,
M.Lang, A.Ochiai, H.Aoki, and T.Suzuki, Phys. Rev. Lett. 86, 2439 (2001).

\bibitem{Affleck}  I.Affleck and M.Oshikawa, Phys.Rev.B 60, 1038 (1999).

\bibitem{Lou}  J.Lou, S.Qin, C.Chen, Z.Su and L.Yu, Phys.Rev.B 65, 064420
(2002).

\bibitem{Ess}  F.H.L.Essler, A.Furusaki and T.Hikihara, Phys.Rev.B 68,
064410 (2003).

\bibitem{kenz}  M.Kenzelmann, R.Coldea, D.A.Tennant, D.Visser, M.Hofmann,
P.Smeibidl, and Z.Tylczynski, Phys. Rev. B 65, 144432 (2002).

\bibitem{kufo}  G.Uimin, Y.Kudasov, P.Fulde and A.A.Ovchinnikov,
Eur.Phys.J.B 16, 241 (2000).

\bibitem{dutta}  A.Dutta, D.Sen, Phys. Rev. B 67, 094435 (2002).

\bibitem{Yang}  C.N.Yang and C.P.Yang, Phys.Rev.B 150, 321, 327 (1966).

\bibitem{mori}  S.Mori, J.-J.Kim and I.Harada, J.Phys.Soc.Jpn 64, 3409
(1995).

\bibitem{hieda}  Y.Hieida, K.Okunishi and Y.Akutsu, Phys.Rev.B 64, 224422
(2001).

\bibitem{XXZhx}  D.V.Dmitriev, V.Ya.Krivnov, A.A.Ovchinnikov and A.Langari,
JETP 95, 538 (2002); D.V.Dmitriev, V.Ya.Krivnov and A.A.Ovchinnikov, Phys.
Rev. B 65, 172409 (2002).

\bibitem{Essler}  J.-S.Caux, F.H.L.Essler, U.Low, Phys.Rev.B 68, 134431
(2003).

\bibitem{sen}  P.Sen, Phys. Rev. E 63, 16112 (2001).

\bibitem{ZZhxhz}  A.A.Ovchinnikov, D.V.Dmitriev, V.Ya.Krivnov,
V.O.Cheranovskii, Phys. Rev. B 68, 214406 (2003).

\bibitem{classical}  J.Kurmann, H.Tomas and G.Muller, Physica A 112, 235
(1982); G.Muller and R.E.Shrock, Phys.Rev.B 32, 5845 (1985).

\bibitem{McCoy}  E.Barouch and B.M.McCoy, Phys. Rev. A 3, 786 (1971).

\bibitem{white}  S.R.White, Phys. Rev. B 48, 10345 (1993).

\bibitem{LSM}  E.Lieb, T.Schultz, D.Mattis, Ann. Phys. (N.Y.) 16, 407 (1961).

\bibitem{Pfeuty}  P.Pfeuty, Ann. Phys. (N.Y.) 57, 79 (1970).

\bibitem{domb}  C.Domb, Adv.Phys. 9, 149 (1960).

\bibitem{Gaudin}  J.des Cloizeaux and M.Gaudin, J. Math. Phys. 7, 1384
(1966).

\bibitem{book}  A.O.Gogolin, A.A.Nersesyan, A.M.Tsvelik, \textit{%
Bosonization and Strongly Correlated Systems} (Cambridge University Press,
Cambridge, 1998).

\bibitem{Hikihara}  T.Hikihara and A.Furusaki, Phys.Rev. B69, 064427 (2004).

\bibitem{BIK}  N.M.Bogoliubov, A.G.Izergin and V.E.Korepin, Nucl.Phys.B 275,
687 (1986).

\bibitem{Cabra}  D.C.Cabra, A.Honecker, P.Pujol, Phys.Rev.B 58, 6241 (1998).

\bibitem{Nersesyan}  A.A.Nersesyan, A.Luther and F.V.Kusmartsev, Phys.Lett.A
176, 363 (1993).

\bibitem{Bocquet}  M.Bocquet, F.H.L.Essler, A.M.Tsvelik and A.O.Gogolin,
Phys.Rev.B 64, 094425 (2001).

\bibitem{Giamarchi}  T.Giamarchi and H.J.Schulz, J. de Physique (Paris) 49, 819
(1988).

\bibitem{Baxter}  R.J.Baxter, Phys. Rev. Lett. 26, 834 (1971); R.J.Baxter,
Ann. Phys. 70, 323 (1972).

\bibitem{Alkaraz}  F.C.Alcaraz and A.L.Malvezzi, J.Phys.A 28, 1521 (1995);
M.Tsukano, K.Nomura, J.Phys.Soc.Jpn.67, 302 (1998).

\bibitem{Q1D}  D.V.Dmitriev and V.Ya.Krivnov, cond-mat/0407203.

\end{thebibliography}
\end{document}